\documentclass[%
 amsmath,amssymb,
preprint,%
]{revtex4-1}

\usepackage{graphicx}
\usepackage{dcolumn}
\usepackage{bm}

\usepackage{cases}
\usepackage{color}

\usepackage[utf8]{inputenc}
\usepackage[T1]{fontenc}
\usepackage{mathptmx}

\begin{document}

\preprint{AIP/123-QED}
\title[]{Identification of wave breaking from nearshore wave-by-wave records}
\author{K. Holand}
\affiliation{Department of Mathematics, University of Bergen, PO Box 7800, 5020 Bergen, Norway}
\author{H. Kalisch}
\email{Henrik.Kalisch@uib.edu}
\affiliation{Department of Mathematics, University of Bergen, PO Box 7800, 5020 Bergen, Norway}
\author{M. Bj\o rnestad}
\affiliation{The Norwegian Meteorological Institute, All\'{e}gaten 70, 5007 Bergen, Norway}
\author{M. Stre\ss er}
\affiliation{Institute of Coastal Ocean Dynamics, Helmholtz-Zentrum Hereon, Geesthacht, Germany}
\author{M. Buckley}
\affiliation{Institute of Coastal Ocean Dynamics, Helmholtz-Zentrum Hereon, Geesthacht, Germany}
\author{J. Horstmann}
\affiliation{Institute of Coastal Ocean Dynamics, Helmholtz-Zentrum Hereon, Geesthacht, Germany}
\author{V. Roeber}%
\affiliation{Universit\'{e} de Pau et des Pays de l'Adour, 
chair HPC-Waves, SIAME, Anglet, France}
\author{R. Carrasco-Alvarez}
\affiliation{Institute of Coastal Ocean Dynamics, Helmholtz-Zentrum Hereon, Geesthacht, Germany}
\author{M. Cysewski}
\affiliation{Institute of Coastal Ocean Dynamics, Helmholtz-Zentrum Hereon, Geesthacht, Germany}
\author{H.G. Fr\o ysa}
\affiliation{Aqua Kompetanse AS, Havbruksparken, Storlavika 7, 7770 Flatanger, Norway}



\date{\today}

\begin{abstract}
Using data from a recent field campaign, we evaluate several breaking criteria
with the goal of assessing the accuracy of these criteria in wave breaking detection.
Two new criteria are also evaluated.
An integral parameter is defined in terms of temporal wave trough area, and a differential
parameter is defined in terms of maximum steepness of the crest front period. 
The criteria tested here are based solely on sea surface elevation derived from standard pressure
gauge records. They identify breaking and non-breaking waves 
with an accuracy between $84 \% $ and $89 \%$ based on the examined field data.
\end{abstract}

\maketitle

\section{\label{sec:level1}Introduction\protect}
Wave breaking is the dominant mechanism of energy dissipation for surface waves in the oceans,
and significant efforts have been made in the past decades to understand various aspects of 
breaking waves both in the coastal ocean and in the open sea \cite{babanin2011breaking}. 
After energy is transmitted from wind to waves during wave generation, waves can traverse
vast distances in the world's oceans, eventually arriving at distant shores.
As waves approach the beach, they tend to increase in height, steepen and eventually break near the beach.
Depending on the beach slope and waveheight, this breaking can take a variety of shapes, and breaking
waves on beaches were classified into spilling, plunging, collapsing and surging \cite{galvin1968breaker}.
Due to its ubiquitous nature and large impact on surfzone dynamics, 
the understanding of breaking waves in shallow water is one of the most important aspects of coastal
wave modeling and the design of coastal structures. Indeed, breaking waves have a major impact on 
sediment transport, beach erosion and exchange of nutrients and other
suspended particles between the surfzone and the inner shelf 
\cite{peregrine1983breaking,davidson2019introduction},
and are also the driving force for the development of surfzone circulation patterns 
\cite{davidson2019introduction,scott2014controls}.

In spite of the prominent role of wave breaking in the study of ocean waves, 
it is one of the least understood ocean surface processes 
\cite{babanin2011breaking, holthuijsen2010waves, toffoli2010maximum}.
As explained in \cite{liberzon2019detection}, one of the main obstacles to
advancing our understanding of wave breaking is the lack of a practical method
for the detection of wave breaking. 
It is generally understood that a wave breaking event commences when the horizontal velocity
of fluid particles near the wavecrest reach the same value as the 
wave velocity \cite{peregrine1983breaking,thornton1983transformation},
and expunged water particles slide down the wavefront in a spilling breaker, 
or the particle velocity eventually exceeds the crest velocity as water is rushed forward 
in an evolving jet \cite{massel2007ocean,banner1993wave,chang1998velocity,kimmoun2007particle,lubin2019discussion}.
So while the start of a breaking event may be defined as above, it is unclear whether such a point
can actually be pinpointed in practice, especially in the case of incomplete information
such as is often the case in field situations which is the main focus of the present work.

Indeed, the definition of breaking onset given above depends on the knowledge of particle velocities 
which are generally difficult to measure in field situations.
As a consequence, indirect methods have been developed to detect wave breaking. In fact, a variety of 
wave breaking criteria based on wave properties such as wave steepness and asymmetry 
have been proposed. In the present note, we analyze recent field measurements
\cite{bjornestad2021lagrangian} in the context of some of the existing breaking criteria
based on wave geometry in order to determine which will work best as a
diagnostic for breaking detection.
The criteria tested include the traditional waveheight to depth threshold,
a number of different wave steepness measures as well as a new criterion
based on an integral of the wave signal.
It is found that the new criterion gives the best overall accuracy, but
all criteria give acceptable levels of accuracy for determining whether
a wave is breaking or not. 
\begin{figure*}
\includegraphics[width=0.56\textwidth]{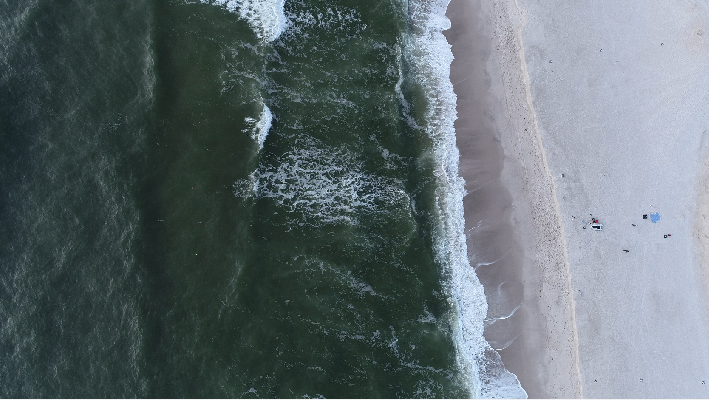}~~
\includegraphics[width=0.42\textwidth]{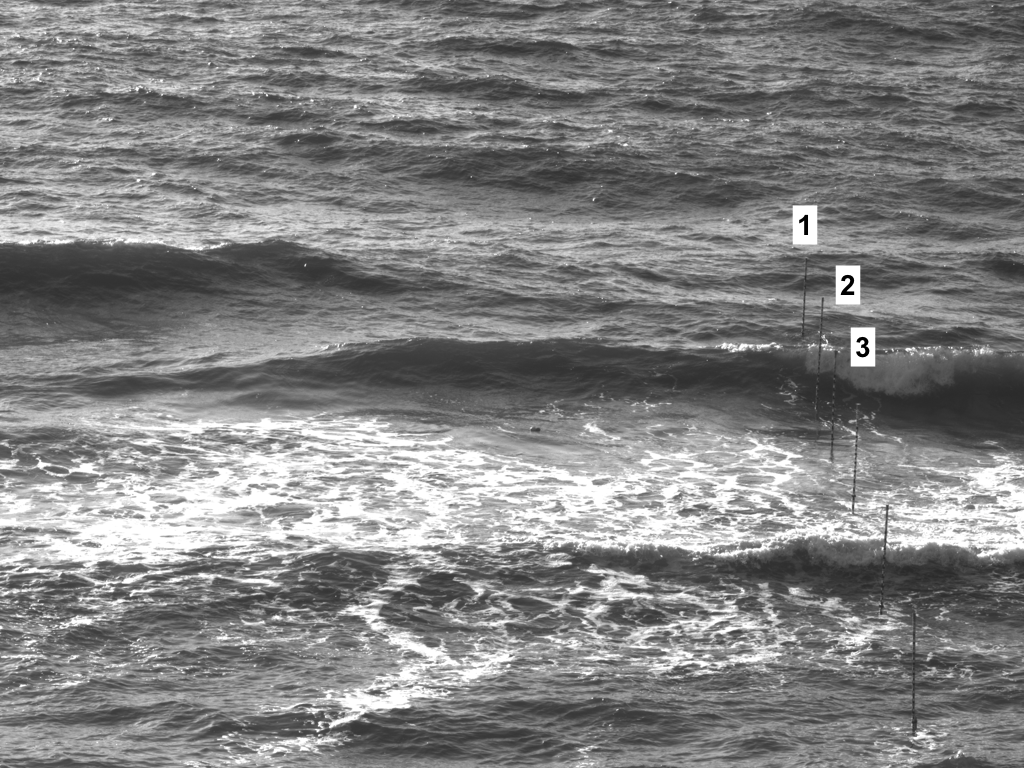}
\includegraphics[width=0.566\textwidth]{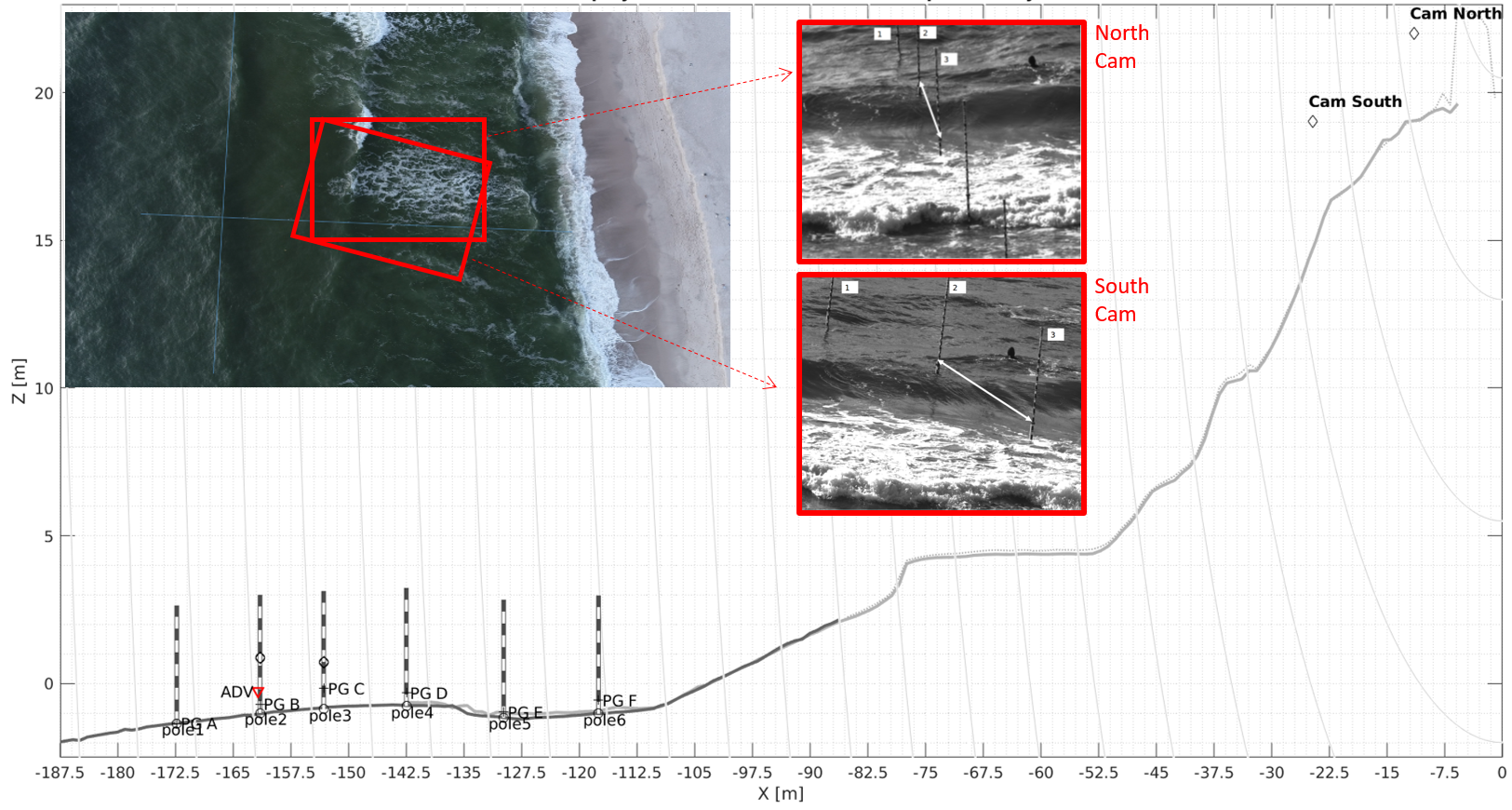}~
\includegraphics[width=0.42\textwidth]{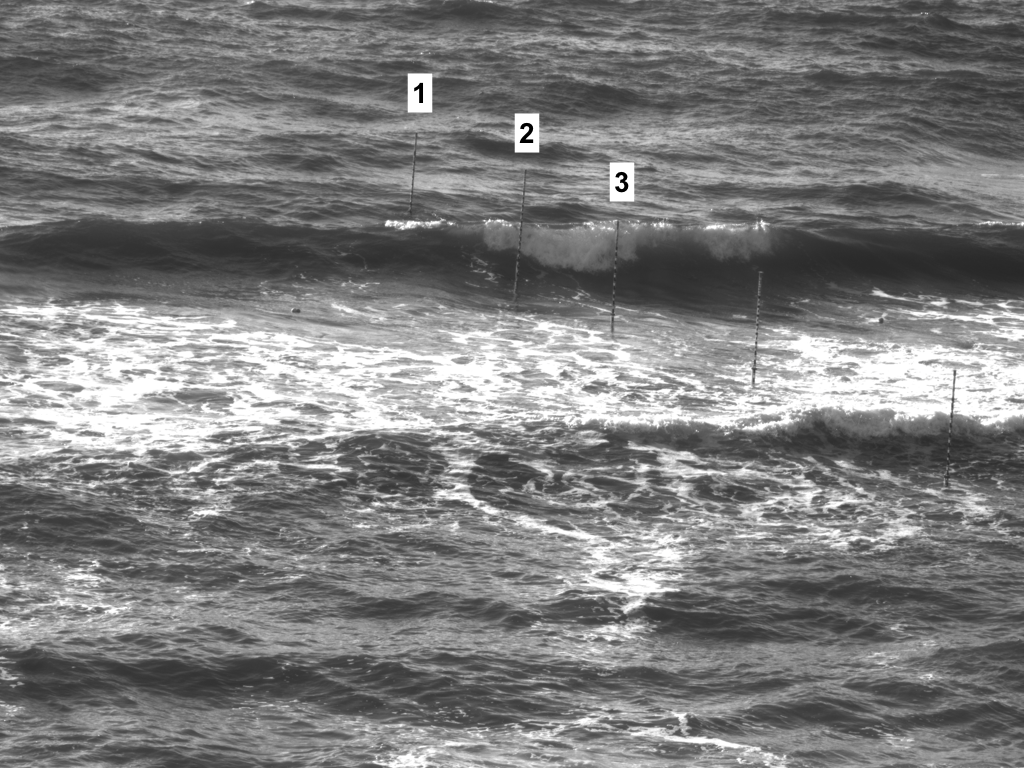}
\caption{\label{fig1} Experimental setup: 
The upper left panel shows an aerial overview 
of the experimental site.
The lower left panel shows the bathymetry, the arrangement of
the poles and the Field of View (FOV)
of the cameras. 
The right panels shows a wider view of the poles
for North Cam (upper) and South Cam (lower).
The pressure signals used here are taken
from the pressure gauges located at the bottom of poles $1$, $2$ and $3$.}
\end{figure*}
\section{Breaking criteria}
Generally, there are three types of criteria used to determine the onset of wave breaking 
(for an in-depth overview, see for example \cite{wu2002breaking,perlin2013breaking} and references therein).
Geometric criteria predict wave breaking using the shape and
more specifically the steepness and asymmetry of the free surface.
Kinematic criteria probe for the violation of the kinematic free surface condition, 
essentially whether stagnation points appear at or near the wavecrest.
Recent works have verified the accuracy of the kinematic criterion,
in particular in shallow water situations \cite{itay2017lagrangian,hatland2019wave},
but if the kinematic criterion is to be used in a practical situation,
estimates of phase or crest velocity have to be provided 
\cite{stansell2002experimental, postacchini2014wave}.
Dynamic criteria are based either on accelerations exceeding
some multiple of the gravitational acceleration \cite{phillips1958equilibrium,bridges2009wave},
or based on relations between energy flux and energy density 
\cite{song2002determining,barthelemy2018unified}.
\begin{figure}[h]
\includegraphics[width=0.48\textwidth]{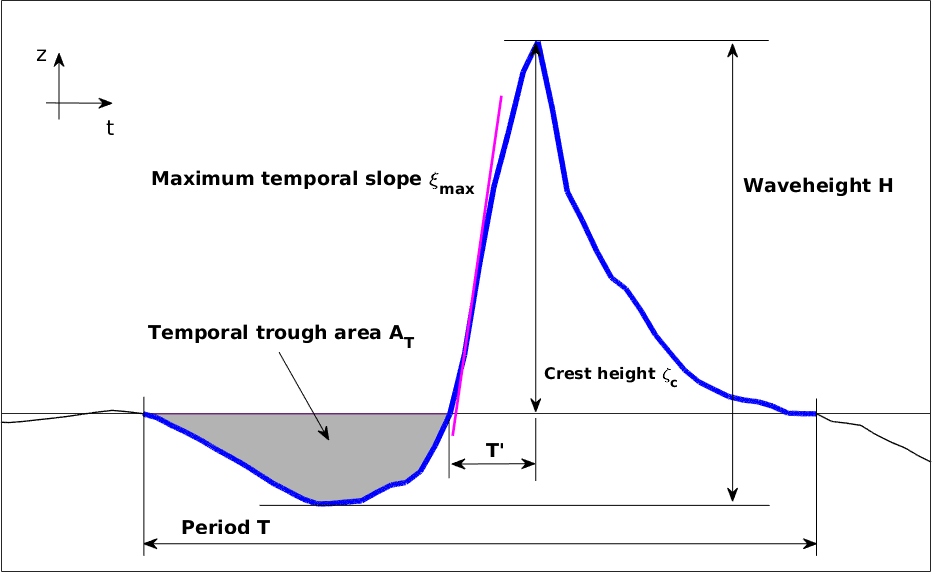}
\caption{\label{fig2} Definition sketch of wave parameters used here.
Waveheight $H$, crest height $\zeta_c$, wave period $T$, wave front period $T'$,
temporal trough area $A_T$ (units: meters$\cdot$seconds) and maximum 
temporal slope $\xi_{\mathrm{max}}$ (units: meters / seconds).
}
\end{figure}
In fact, there are several physical mechanisms which can lead to wave breaking,
for example crest instabilities in deep water \cite{longuet1997crest},
bottom forcing in coastal regions \cite{kirby2019short, briganti2005boussinesq},
wind forcing \cite{babanin2011breaking} and forced discharge \cite{favre1935etude}.
In general, one should distinguish between deep-water wave breaking
(i.e. in the open ocean or on a lake, far from the shore) and shallow-water breaking,
i.e. depth-induced breaking near the shore.
\begin{table}[b!]
\caption{\label{Table1} Wave breaking indicators. The indicator $\kappa$
is defined in  \eqref{KAP}. The indicator $\xi_{\mathrm{max}}$ is defined
in \eqref{MAX}. The parameter $\zeta_c$ is the crest height, $T$ is the
wave period, $g$ is the gravitational acceleration, $T'$ is the
wave front period, $H$ is the waveheight and $h_0$ is the depth.}
\begin{ruledtabular}
\begin{tabular}{l c r r}
Criterion &    Indicator &        Units\\
    \hline
   Integral criterion  &  $\kappa$   &  \\
    Maximum steepness & $\xi_{{\mathrm max}}$  & m/s  \\
    Steepness  I  &  $\zeta_c / T'$ &  m/s\\
    Steepness  II & $\frac{\zeta_{c}}{(g/2 \pi) T \cdot T'} $  &  \\
    Steepness  III  &  $H / T$  &  m/s\\  
   Waveheight/depth & $ \gamma = H/h_0$  &   \\ 
\end{tabular}
\end{ruledtabular}
\end{table}
\newpage
\begin{figure*}
\includegraphics[width=0.64\textwidth]{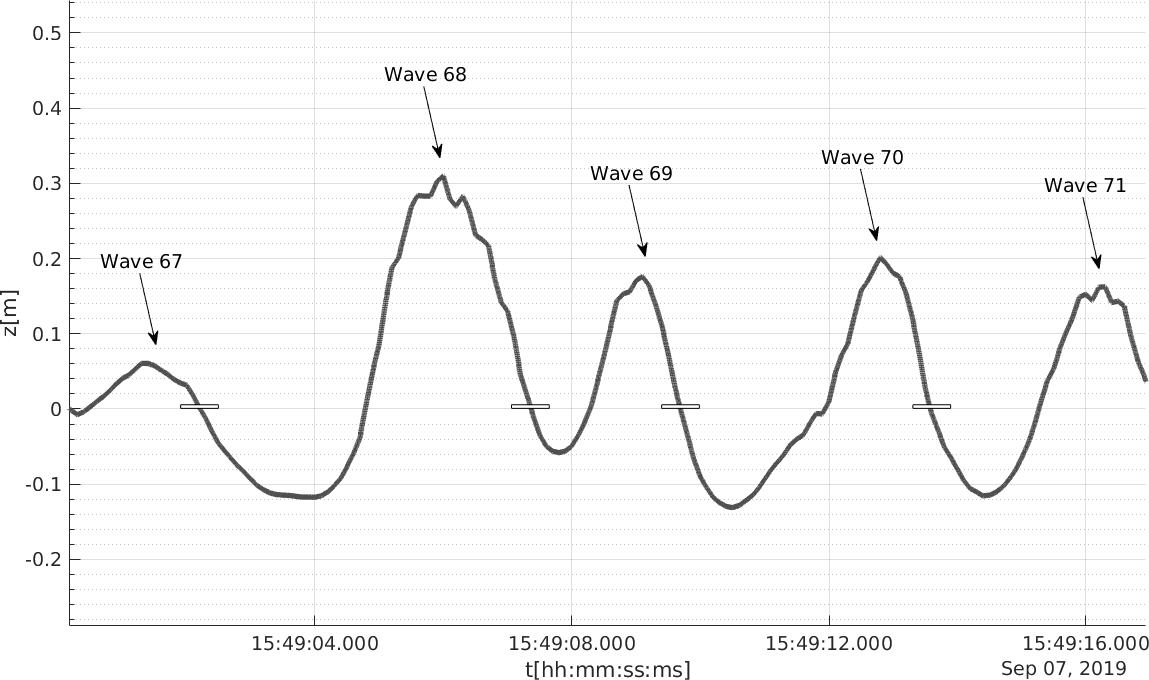}
\caption{\label{fig3} 
Segmentation of the wave record. A zero down-crossing segmentation is applied
to each wave record. In this figure, five waves in the record Cv48 at Pole 2
are shown (Wave 67 through Wave 71). The white bar designates the demarcation
of two different wave events.
For each wave in each record the basic parameters indicated in Figure 2 are found,
and the six quantities delineated in Table 1 are computed.}
\end{figure*}

Studies of wave breaking in shallow water have mostly focused on the breaker height 
following the pioneering work of McCowan \cite{mccowan1894xxxix}
and later Munk \cite{munk1949solitary}
where the limiting relative waveheight for breaking solitary waves was found 
in terms of the waveheight to depth ratio $H/h_0$.
The critical value of this ratio depends on a number of factors, and even for a flat bed,
it is not entirely clear what the critical value should be \cite{massel1996largest}.
In fact, many works have focused on empirical fits of the so-called breaker index
the critical value of $\gamma$ at which waves are expected to break.
These studies are based on a number of dedicated laboratory and field studies with various bed slopes.
For example, Madsen \cite{madsen1976wave} defines a breaker index $\gamma_b = 0.72(1+6.4m)$, where
$m$ is the bed slope, and Battjes \cite{battjes1974surf} defines  $\gamma_b = 1.062 + 0.137 \log (\xi_0)$
in terms of the surf similarity parameter $\xi_0 = m \sqrt{L_0/H_0}$,
where $H_0$ is the offshore waveheight and $L_0$ is the offshore wavelength.
An overview over much of the existing literature can be found in \cite{robertson2013breaking}.

The main purpose of the present work is to test a number of 
wave breaking criteria as a simple diagnostic for deciding 
whether an individual wave in a given record is breaking or not.
The diagnostic is based only on time series data of the free surface elevation. 
This time series could be obtained from a wave gauge or from a pressure sensor mounted 
in the fluid column or near the fluid bed.  
In this situation, the class of criteria based on wave shape appear to be most
expedient. In some works which analyze data from laboratory experiments, 
the Phase-Time Method (PTM)
\cite{huang1992local, griffin1996kinematic, stansell2002experimental,itay2017lagrangian},
or the wavelet method \cite{longo2003turbulence,longo2009vorticity,massel2007ocean} is used.
Such an analysis would have to use the Hilbert transform 
to estimate phase and particle velocities \cite{massel2007ocean} and would be inapplicable
to field situations unless a special setup were to be used.
In the present case, we focus on situations where common devices such as pressure gauges
or single wave gauges are used, and the diagnostic should therefore use methods
that require minimal postprocessing.

The criteria tested here are summarized in Table \ref{Table1}. 
We test the traditional waveheight / depth criterion,
as well as three different steepness criteria.
For a given wave record, a wave-by-wave segmentation is applied,
and each wave is assigned a number (see Figure \ref{fig3}). 
For each numbered wave, the basic quantities
waveheight $H$, wave period $T$ and crest height $\zeta_c$ are found numerically
(see Figure \ref{fig2}). In addition, the wave front period $T'$, i.e. the time
between a zero-upcrossing until the wave crest is reached is found.
From these quantities, the waveheight/depth ratio $\gamma = H/h_0$,
and the three steepness parameters $\zeta_c/T$, 
$\frac{\zeta_{c}}{(g/2 \pi) T \cdot T'} $ and  $H / T$ are computed for
each wave in a given record.

In addition, we define a new parameter based on the size of
the trough preceding a wave crest. 
This parameter is based on the observation that an extensive
wave trough is often preceding a breaking wave.
Hand in hand with a large trough goes a large steepness of the wave front, not necessarily
as defined by the usual measures, but rather locally, so we also defined a new steepness
criterion based on the
maximum steepness (in terms of the temporal slope) of the wave front.
We thus define wave breaking diagnostics on an integral measure, the
size of the preceding trough (called {\em temporal trough area}  $A_T$)
and a differential measure: the maximum slope of the crest front $\xi_{max}$.
The exact definitions are as follows. We define the non-dimensional quantity
\begin{equation}
\label{KAP}
\kappa = \frac{H^2 \cdot A_T}{T \cdot h_0^3},
\end{equation}
where $H$ is the waveheight, $A_T$ is the temporal trough area (units m$\cdot$s),
$T$ is the wave-by-wave period and $h_0$ is the fluid depth.
The temporal trough area is defined by
$A_T = \int_{t_{\mathrm{down}}}^{t_{\mathrm{up}}} | \eta(t) | \, dt$,
where $_{t_{\mathrm{down}}}$ denotes
the time of zero-down crossing defining the starting point of the the
wave,
and  $t_{\mathrm{up}}$ denotes the up-crossing time immediately following  $t_{\mathrm{down}}$
(see Figure \ref{fig2}) for a definition sketch.
The maximum temporal slope is defined as 
\begin{equation}
\label{MAX}
\xi_{\mathrm{max}} =  \nu_S \cdot \max_{t_i} \left\{ \eta(t_i) - \eta(t_{i-1}) \right\},
\end{equation}
where $\nu_S$ is the sampling frequency and $\eta(t_i)$ are free surface records between
$t_{\mathrm{up}}$ and $t_{\mathrm{down}}$.
While the integral measure may have the advantage of being more stable
due to an inclusion of the signal history, both measures work almost equally
well for the wave records considered here.

\section{Field measurements}
The measurements described here were obtained from a campaign that took place during 4-8 September 2019,
on the western coast of Sylt, an island off the German North Sea Coast
using a combination of both in situ and remote sensing measurement systems. 

A long-range, high resolution four-camera stereo imaging system was specifically developed for this study.
Two pairs of 5MP, global shutter CMOS digital cameras (Victorem 51B163-CX, IO Industries) were each fitted 
with Canon 50 mm and 400 mm lenses, respectively. The two camera pairs were placed on the ridge overlooking the beach, 
at a distance of $40$m from one another. The four cameras were focused on a portion 
of water surface within the surf zone, located at an approximately distance of $150$m 
from the cameras. A sketch of the instrument setup is provided in Figure \ref{fig1}.

Six graduated aluminum poles were jetted into the sand of an intertidal sandbar at low tide.  
The array of poles was aligned so as to be approximately perpendicular to the crests of incoming 
waves. The most seaward pole (Pole 1) was about $80$ m from the shore,
and the closest pole (Pole 6) was about $20$ m from the shoreline. 
At the base of each pole, a pressure gauge measured absolute pressure at $10$ Hz sampling frequency. 
The recorded pressure signal was subdivided into $10$ minute data bursts and then
transformed to surface excursion using the nonlinear method encapsulated in eq. (13) in \cite{bonneton2018nonlinear}.
This method has been found to be quite accurate, with the highest error of $\sim 7 \%$ at the wavecrest
(see also \cite{mouragues2019field}). 
Since the graduated poles were within the field of view of the stereo cameras (acquiring at $30$ frames/second), 
these were also used as optical wave gauges in order to verify the nonlinear re-construction of
the free surface.
\begin{figure*}
\includegraphics[width=0.46\textwidth]{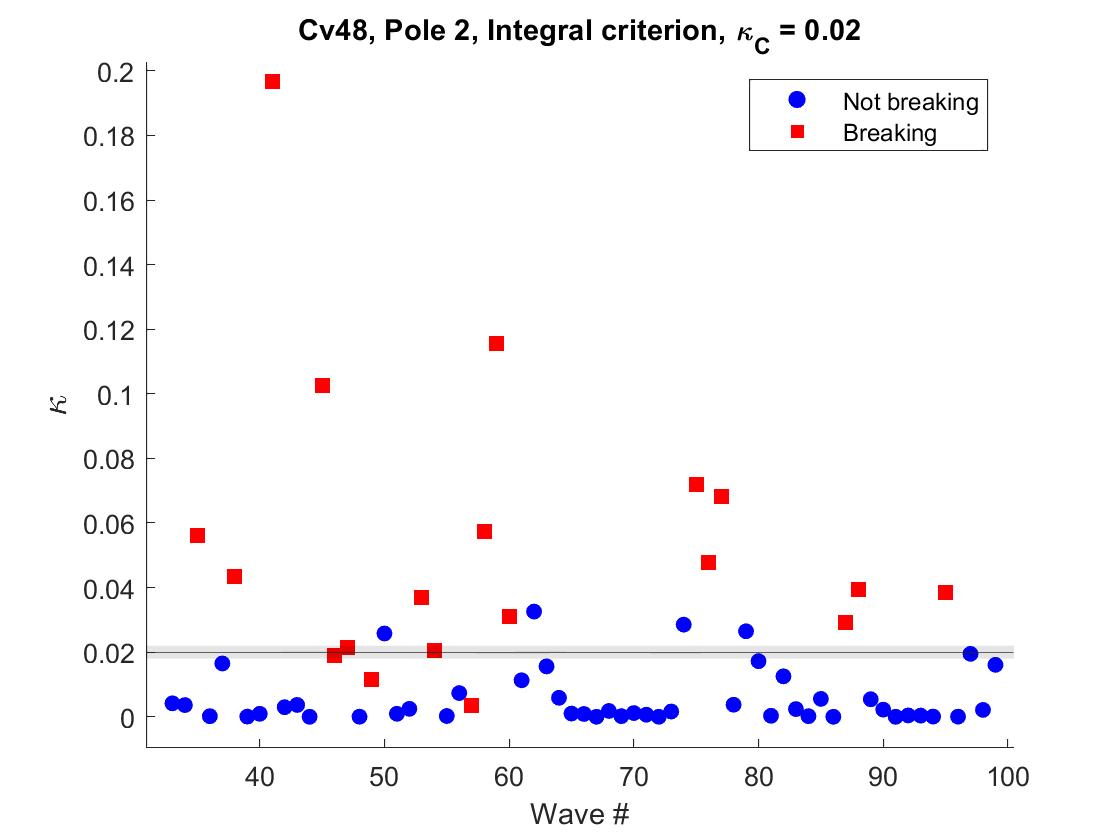}
\includegraphics[width=0.46\textwidth]{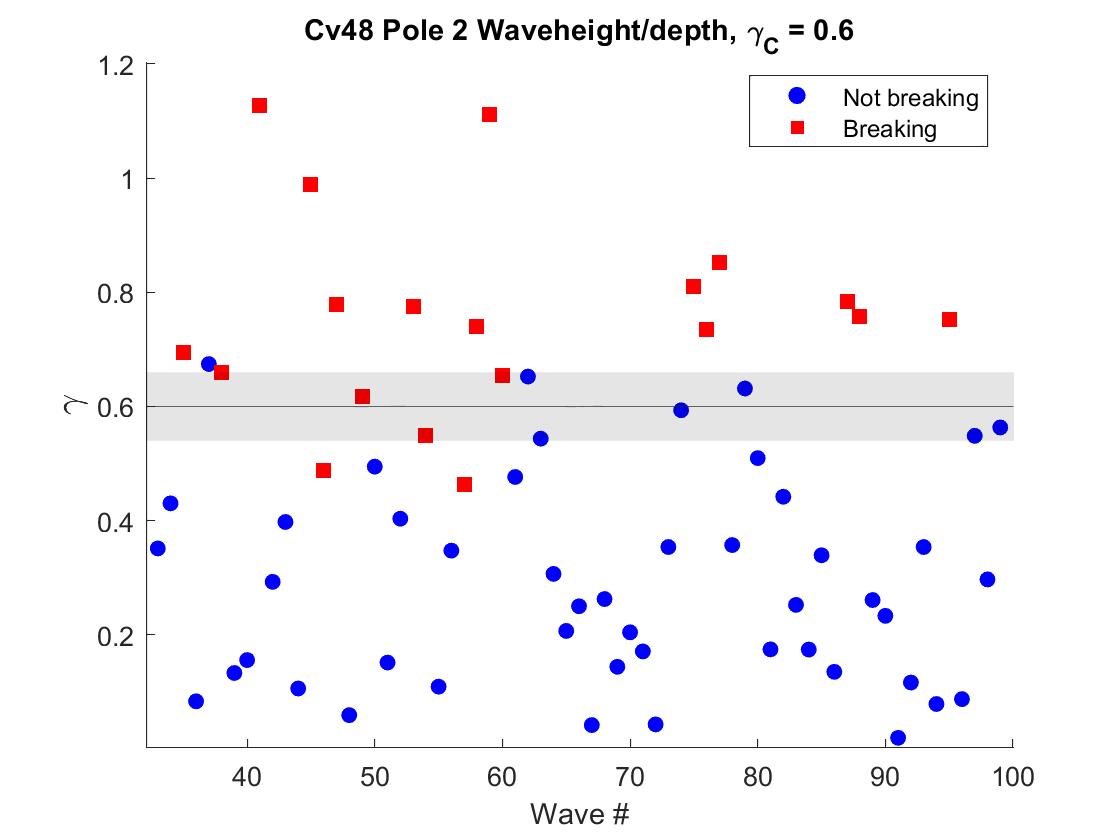}
\caption{\label{fig4} Graphical representation of the identification of breaking and non-breaking
waves for dataset Cv48 at Pole 2: The left panel shows evaluation of the integral criterion 
for  breaking waves (red) and non-breaking waves (blue).
The right panel shows the evaluation of the waveheight / depth criterion for 
breaking waves (red) and non-breaking waves (blue). 
The gray shaded area represents a $10\%$ tolerance band for the critical value
to take account of various errors in the measurements and imperfections in the data analysis
such as the free surface reconstruction.}
\end{figure*}

\begin{figure}
\includegraphics[width=0.48\textwidth]{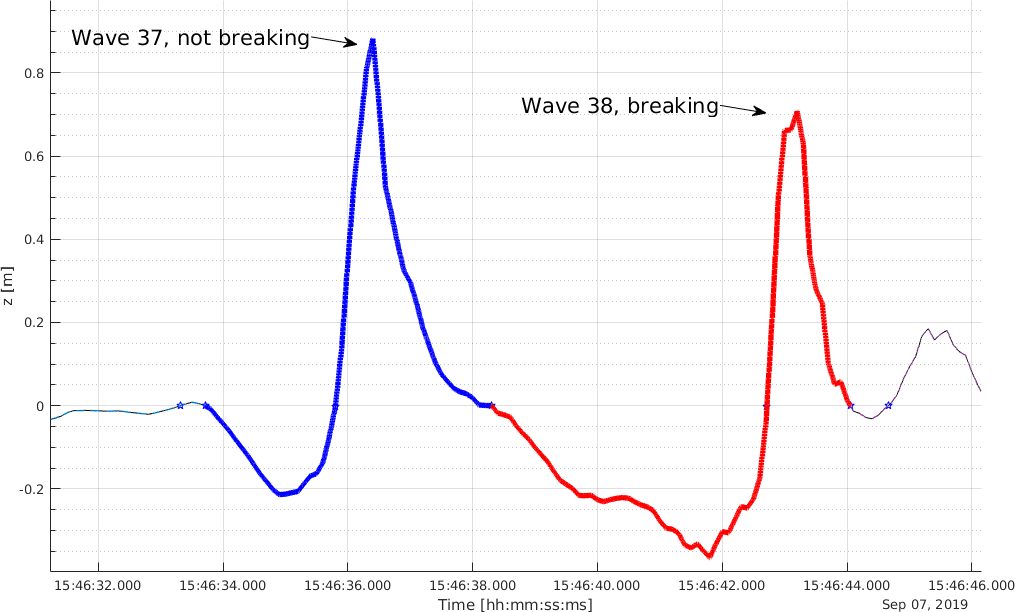}
\caption{\label{fig5} 
Excerpt from data set Cv48 showing two wave profiles. The first wave (wave 37, shown in blue) 
is not breaking, while the second wave (wave 38, shown in red) is breaking. All traditional
diagnostics based on wave shape fail to classify these waves accurately.}
\end{figure}
\begin{figure*}
\includegraphics[angle = 0,width=0.96\textwidth]{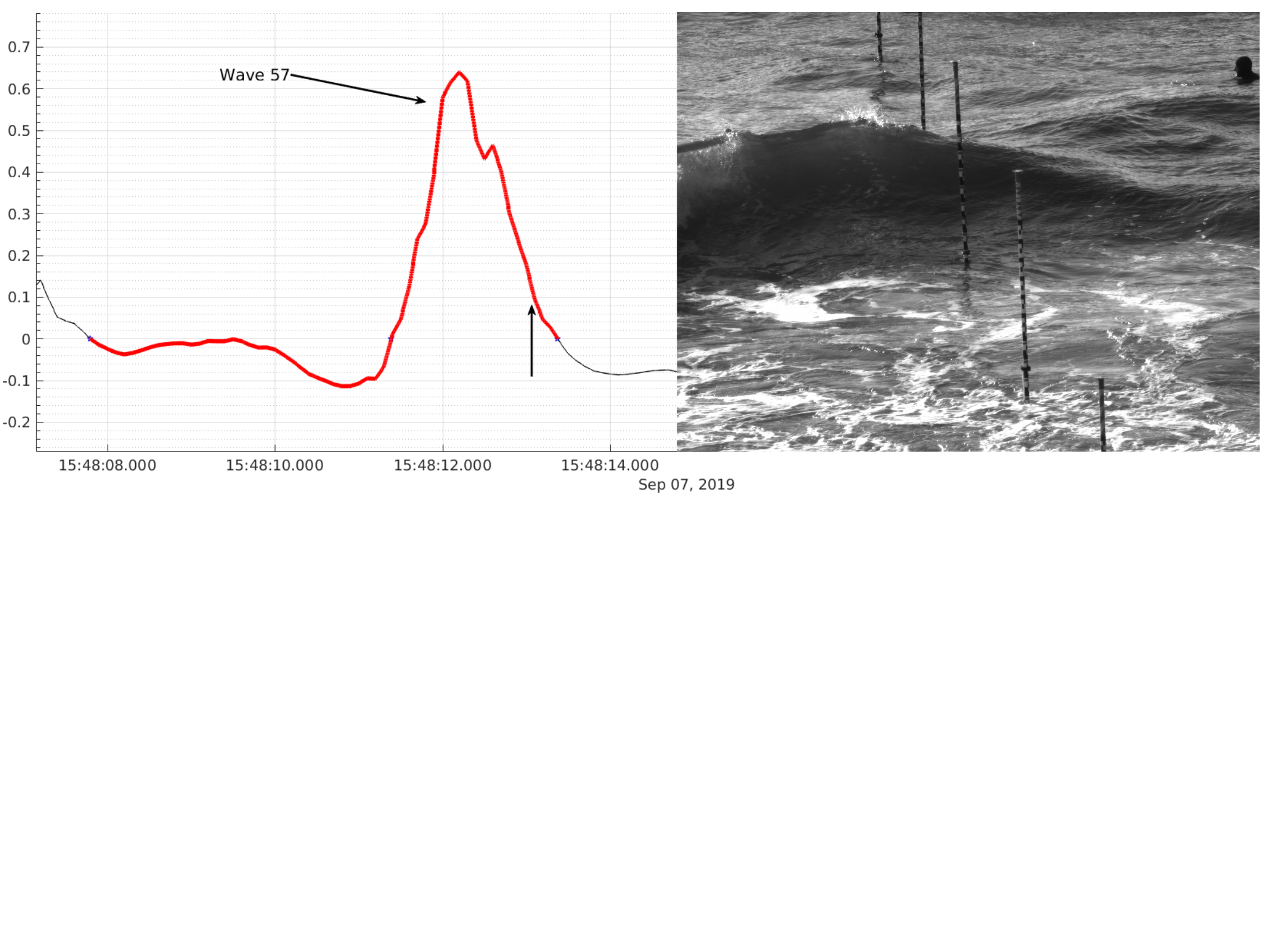}
\vskip -2.5in
\caption{\label{fig6} 
One of the waves in the record Cv48 (Wave 57) which exhibited breaking at Pole 2, but which did
not trigger either the waveheight/depth or the integral criterion.
As can be seen in the frame from the North Cam,
the reason appears to be that the wave is short-crested and coming in to shore at a slight angle,
so that the correct signal history with regard to wave-breaking prediction is not available at Pole 2.
The time series at pole 2 is shown in the left panel, and a single frame from the North Cam
is shown in the right panel. The vertical arrow in the left panel denotes the time stamp
from the frame in the right panel.}
\end{figure*}
\begin{figure*}
\includegraphics[angle = 0,width=0.96\textwidth]{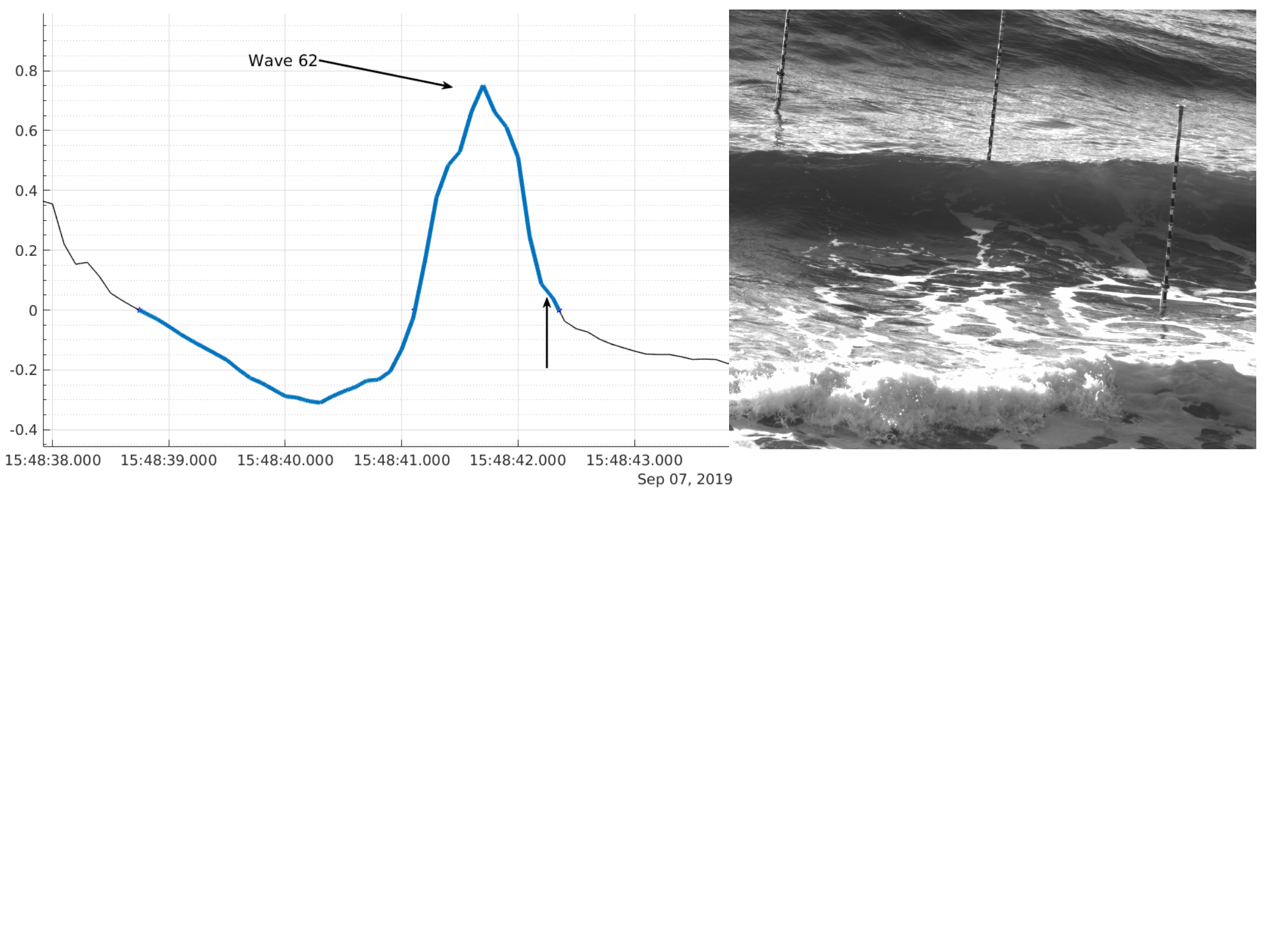}
\vskip -2.5in
\caption{\label{fig7} 
One of the waves in the record Cv48 (Wave 62) which did not break at Pole 2, but which did
trigger both criteria shown in Figure \ref{fig4}. The reason why breaking was retarded is not clear.
Wind effects are a possibility.
The time series at pole 2 is shown in the left panel, and a single frame from the North Cam
is shown in the right panel. The vertical arrow in the left panel denotes the time stamp
from the frame in the right panel.}
\end{figure*}
\begin{figure*}
\includegraphics[width=0.96\textwidth]{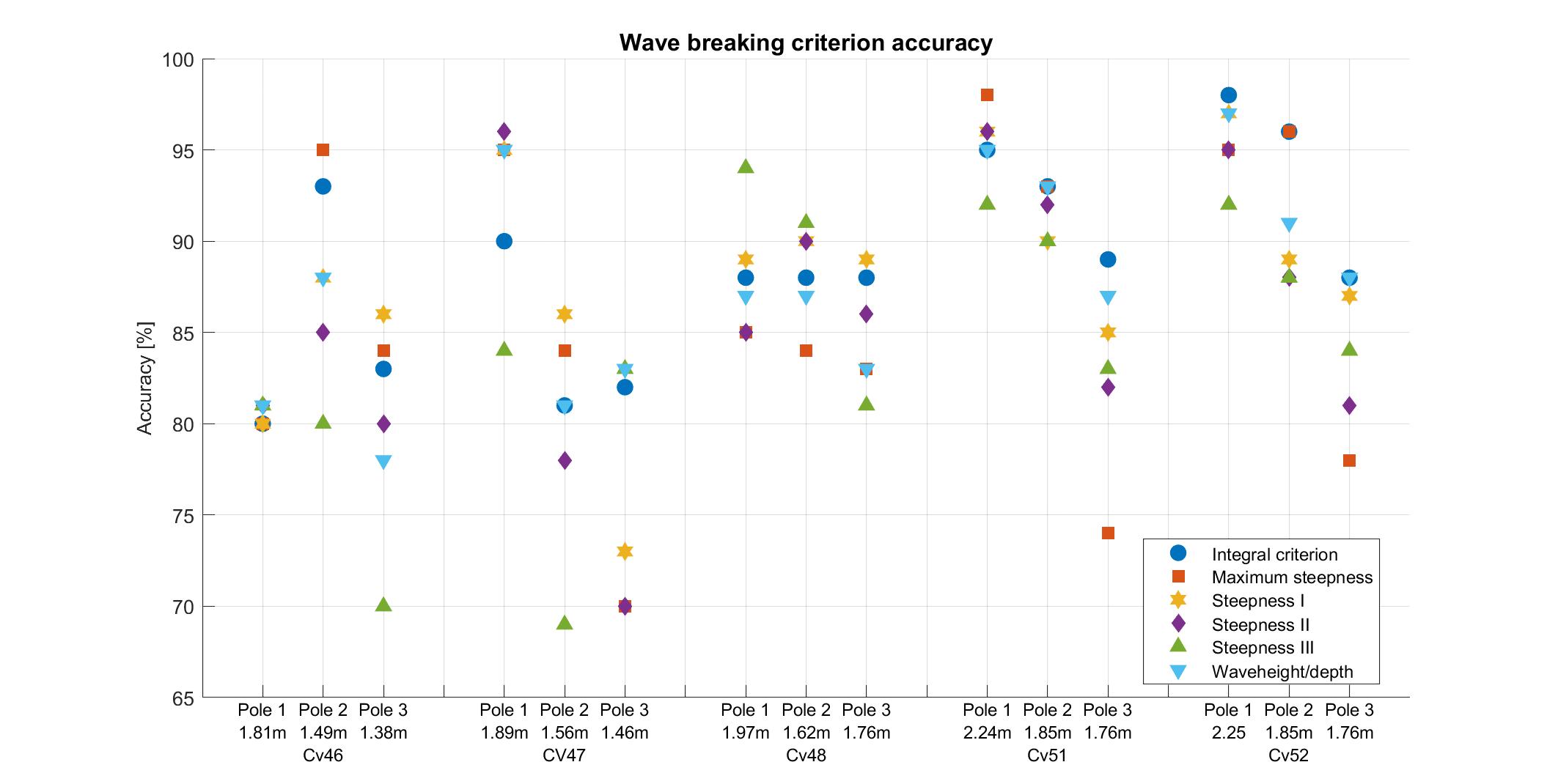}
\caption{\label{fig8} Accuracy of the six different criteria in detecting breaking waves
across all $15$ datasets.}
\end{figure*}
\section{Data analysis}
The data consists of pressure data and video frames of the sea surface at a shore in Sylt, Germany
recorded in the period between 15:13:00 and 17:18:59 UTC on September 7th, 2019. 
In total, $903$ wave events distributed over five data sets
(datasets Cv46, Cv47, Cv48, Cv51 and Cv52) were analyzed.
The waves were collocated at the first three poles (Pole $1$, Pole $2$ and Pole $3$)
with the corresponding time series from the pressure gauge
records.
The free surface elevation is reconstructed from the pressure data using the
method explained in \cite{bonneton2018nonlinear}.
The sea surface time series is adjusted for tidal effects, 
and the approximate depth during one wave record is obtained by averaging
over the entire $10$-minute record.

Wave conditions were monitored at an offshore buoy located in about $10$m water depth.
Conditions for significant waveheight were in the range $0.9 - 1$m, peak period
was in the range $6.25 - 6.7$s, and peak direction was in the range $270-289^{\circ}$.
The overarching aim here is to find a criterion for determining whether a given wave 
in the record is breaking or not, based solely on the free surface time series derived from the pressure data.
The visual images are only used for verification of the diagnostic.

Overall, at Pole 1, $20$ out of $293$, or $7\%$ of waves are breaking.
At Pole 2, $83$ out of $300$, or $28\%$ of waves are
actively breaking,
and $75$ out of $310$ or $24\%$ of waves
are actively breaking at Pole 3.
The water height usually decreases from Pole $1$ to Pole $3$ during the period of measurements 
which explains the different percentages of breaking waves for the different locations.
At Pole 4 almost all waves have broken or are actively breaking, and at Pole 5 and 6,
almost all waves have broken.

Wave breaking was defined by visual inspection, and a wave was counted as breaking
at a given pole if breaking occurred in the vicinity of the pole. In some cases,
ambiguities occurred, such as breaking of secondary crests riding on top of
the main wave. If such an event 
was intermittent, lasting less than $1$ second,
this was not counted as a breaking wave.
\begin{table}
\caption{\label{Table2}
Accuracy of the six breaking detection criteria at each of the three poles. The overall
accuracy shown in column $5$ is given with an error which is determined by using a $10\%$ 
error bar for the demarcation of individual wave events as shown in Figure \ref{fig4}.}
\begin{ruledtabular}
\begin{tabular}{l c c c c}
 Criterion  & Pole 1  & Pole 2  & Pole 3 &  Overall accuracy \\
    \hline
     Integral criterion &  $91\%$  & $90\%$ & $87\%$ & $89\%  \pm 1$\% \\ 
     Max. steepness &  $92\%$  & $90\%$ & $77\%$ & $86\%  \pm 1$\% \\ 
     Steepness I &  $93\%$  & $90\%$ & $84\%$ & $87\%  \pm 2$\% \\ 
     Steepness II &  $92\%$  & $87\%$ & $81\%$ & $86\%  \pm 3$\% \\ 
     Steepness III &  $90\%$  & $85\%$ & $84\%$ & $86\%  \pm 2$\% \\ 
     Waveheight/depth  &   $93\%$  & $88\%$ & $84\%$ & $88\%  \pm 1$\% \\ 
\end{tabular}
\end{ruledtabular}
\end{table}

In order to test the criteria under examination here,
critical values for each diagnostic parameter must be found.
The approach taken here was to calibrate the critical value of a
diagnostic parameter using one of the $15$ datasets 
(here we used Cv52 at Pole 2, but any other dataset could have been used).
Once calibrated, the critical value was applied unchanged to the remaining
datasets.

In order to account for the up to $7 \%$ error in the free surface reconstruction
and various other small errors in the measurements, we incorporated a $10 \%$ tolerance
band around the critical value of each diagnostic parameter. As can be clearly see
in Figure \ref{fig4}, the accuracy in terms of share of correctly identified waves
is rather stable with respect to this tolerance. For example, an increase from $10 \%$
to $15 \%$ would result in an increased error of only $1-2 \%$ in the overall accuracy.

Overall, the traditional criteria {\em Steepness I}, {\em Steepness II}, {\em Steepness III}
and {\em Waveheight / depth} with the corresponding formulae
given in Table \ref{Table1} yield acceptable results for wave breaking identification.
The best of these four criteria is the {\em Waveheight / depth} criterion
with overall $88\%$ accuracy across the $903$ events studied here (see Table \ref{Table2}).
For a subset of wave events (Cv 48, Pole 2), the accuracy of the {\em Waveheight / depth} criterion
is depicted in the right panel of Figure \ref{fig4}. The red squares signify waves which
were visually inspected to be breaking at Pole 2 while the blue dots denote waves which
are not breaking. The value of $\gamma$ is indicated on the ordinate.

There are some constellations of waves where all of the traditional criteria
give counter-intuitive results. Consider the two waves from record Cv48
shown in Figure \ref{fig5}.
The wave on the left (Wave $37$) is not breaking (indicated in blue)
while the wave on the right (Wave $38$) is breaking (indicated in red).
For each of the traditional criteria, the value of the corresponding
indicator is higher for Waves $37$ then for Wave $38$.
The decisive property that appears to override all other metrics is the
extensive wave trough preceding Wave $38$. This deep trough essentially
lowers the water depth, so that the succeeding wave crest is high enough
relatively to the lower preceding depth to lead to wave breaking.
This deep trough in combination with a still relatively large crest height
leads to a steep wave front which is most easily detected with a local
measure of steepness.
These observations led us to define the {\em Integral criterion}
(top line in Table \ref{Table1}) and the {\em Maximum steepness criterion}
(second row in Table \ref{Table1}).
As shown in Table \ref{Table2}, the {\em Integral criterion}, represented by
the indicator $\kappa$ defined in \eqref{KAP} gives the highest
overall accuracy, and also works evenly across various observational records.

Each of the six criteria gives some false positives and false negatives.
Two of such are shown in Figure \ref{fig6} and Figure \ref{fig7}.
For the wave shown in Figure \ref{fig6}, it is evident that it is short-crested,
and the immediate elevation history at a single location (in this case Pole 2)
is skewed, and will not allow an accurate classification of the wave with
regards to breaking. The wave shown in Figure \ref{fig7} triggered all breaking criteria,
but did not break until it was too far from Pole 2 to be counted. It is not
immediately obvious what caused the discrepancy.

\section{Discussion and outlook}
In the present work, it has been demonstrated that breaking waves can be detected
from nearshore wave-by-wave records with an $84 \%$  to $89 \%$ accuracy, at least based
on the records from recent field measurements examined here (see Figure \ref{fig8}).
Six criteria have been tested, and they all give acceptable results.
A new integral criterion based on trough size of a wave has been put forward.
While the new criterion gives the best overall performance, the improvement
is too small to justify the additional complication of the temporal integration.

The breaking detection tested here works with a single wave gauge or pressure sensor.
Environmental parameters such as precise bathymetry measurements, wind
and current effects have purposely not been taken into account as we were aiming for a simple
diagnostic which should give acceptable results in situations were such data are not available. 
Nevertheless, it would be interesting to test wave breaking detection based on these simple
diagnostics in a controlled environment such as a wave flume or wave basin. Such a study might
also cast more light onto why some false positives appear, for example Wave 62 shown in Figure
\ref{fig7} which triggered all criteria, but did not break close enough to Pole 2 to count
as breaking.

The critical values of each diagnostic parameter was found
using one of the $15$ datasets, and then applied to the remaining records. 
It will be interesting to see whether some of these critical values hold
also in other situations. 
For the critical waveheight-to-depth parameter value $\gamma_c$, 
a rather wide range of values has been suggested \cite{robertson2015remote}
(it appears however that most of the criteria have been validated only
for laboratory data).
Using the Madsen criterion with the bed slope of $\sim 1:50$ at the
experimental site, and the offshore wave conditions given by the
buoy in $10$m depth, a critical value of $\sim 0.81$ is found,
and the Battjes formula yields a critical value of $\sim 0.78$.
Other works \cite{massel1996largest, brun2018convective}
indicate a critical breaker height close to $0.6$
which is similar to the critical value found here during the calibration. 
As indicated already in \cite{massel1996largest}, more field studies
are required in order to draw any conclusions on whether there
is a universally applicable breaker height definition.

Previous measurements and simultaneous visual observation are primarily available
for deep-water situations
(see for example \cite{holthuijsen1986statistics, katsaros1992dependence, babanin2011breaking}).
In \cite{holthuijsen1986statistics}, it is suggested that geometric parameters such as local
asymmetry and steepness cannot be used with confidence to determine whether
a given surface record features a breaking or non-breaking wave.
In contrast, we find that the criteria used here give the correct
determination for close to $90 \%$ of all wave events.
Previous studies successfully applying wave-by-wave properties of wave records
in the context of wave breaking exist \cite{postacchini2014wave, liberzon2019detection},
and partially motivated the current work. 

While the present paper focuses on breaking detection, significant efforts have also been directed
towards predicting wave breaking by identifying the point of breaking inception
\cite{barthelemy2018unified, derakhti2020unified}.
Both methodologies are of importance for numerical ocean modeling.
Breaking detection should be applied for preparing ocean data as input for numerical models
while breaking prediction can be used to understand when numerical dissipation should be used
to simulate wave breaking. In fact, recent works have illuminated the use of 
various wave-breaking criteria in Boussinesq-type models, 
and a number of different approaches have been implemented and tested 
\cite{roeber2010shock,bjorkavaag2011wave,tonelli2011simulation, tissier2012new,bacigaluppi2020implementation}.
While the waveheight-to-depth and steepness criteria have been mostly used as {\em breaking inception} criteria, 
here they have been indicated to work well as {\em detection criteria}.

\begin{acknowledgments}
We thank Jan B\"{o}dewadt and Jurij Stell for technical support. 
We acknowledge funding from the Research Council of Norway under grant no. 239033/F20,
and from Bergen Universitetsfond,

MPB, JH, MS, MC, and RC wish to acknowledge 
from the Coasts in the Changing Earth System (PACES II) program of the Helmholtz Association,
and support from the Deutsche Forschungsgemeinschaft 
(DFG, German Research Foundation, project number 274762653, Collaborative Research Centre TRR 181 
{\em Energy Transfers in Atmosphere and Ocean}). 

Volker Roeber acknowledges financial support from the I-SITE program
{\em Energy \& Environment Solutions} (E2S), the Communaut\'{e} d'Agglom\'{e}ration Pays Basque (CAPB),
and the Communaut\'{e} R\'{e}gion Nouvelle Aquitaine (CRNA)
for the chair position HPC-Waves and support from the European Union's Horizon 2020
research and innovation programme under grant agreement no. 883553. \\
\end{acknowledgments}


\providecommand{\noopsort}[1]{}\providecommand{\singleletter}[1]{#1}%

\end{document}